\documentclass[a4paper,11pt]{article}
\usepackage{graphicx,latexsym,color,amsmath,amsfonts}
\usepackage{cite,citesort,pgfplots,color}

\newcommand{\diff}{\mathop{\mathrm{\mathstrut{d}}}\!}%Diff op.

\newcommand{\eps}{\varepsilon}
\newcommand{\he}{\hat{\boldsymbol{e}}}
\newcommand{\hz}{\hat{\boldsymbol{z}}}

\newcommand{\hht}{\hat{\boldsymbol{h}}_t}

\begin{document}
\section*{}
Disclaimer: This work has been accepted for publication in the IEEE Antenna and Wireless Propagation Letters.  Copyright with IEEE. Personal use of this material is permitted. However, permission to reprint/republish this material for advertising or promotional purposes or for creating new collective works for resale or redistribution to servers or lists, or to reuse any copyrighted component of this work in other works must be obtained from the IEEE. This material is presented to ensure timely dissemination of scholarly and technical work. Copyright and all rights therein are retained by authors or by other copyright holders. All persons copying this information are expected to adhere to the terms and constraints invoked by each author’s copyright. In most cases, these works may not be reposted without the explicit permission of the copyright holder. For more details, see the IEEE Copyright Policy.
\newpage

\title{Array Antenna Limitations}
\author{B.~L.~G.~Jonsson, C.~I.~Kolitsidas {\it student member, IEEE} and N.~Hussain%
\thanks{All authors are with the School of Electrical Engineering, KTH Royal Institute of Technology, SE-100 44 Stockholm, Sweden, email:lars.jonsson@ee.kth.se}}
\maketitle

\begin{abstract}
This letter defines a physical bound based array figure of merit for both single and multi-band array antennas. It provides a measure to compare their performance with respect to return-loss, bandwidth(s), thickness of the array over the ground-plane, and scan-range. The result is based on a sum-rule result of Rozanov-type for linear polarization. For single-band antennas it extends an existing limit for a given fixed scan-angle to include the whole scan-range of the array, as well as the unit-cell structure in the bound. The letter ends with an investigation of the array figure of merit for some wideband and/or wide-scan antennas with linear polarization. We find arrays with a figure of merit $>0.6$ that empirically defines high-performance antennas with respect to this measure.  
\end{abstract}
%\begin{IEEEkeywords}
%Array antenna limitations, antenna theory, electromagnetic analysis, all-in-one arrays, figure of merit, physical bound, bandwidth limitation. 
%\end{IEEEkeywords}

\section{Introduction}
%\IEEEPARstart{B}{roadband} 
Broadband 
arrays compete with multiband arrays in providing communication bandwidth service over a very large frequency band. Factors like array thickness, scan-range and return-loss are important in the choice of base-station antenna elements. It is well known that the cellular phone spectrum distribution and its variations with country and network technology poses challenges on the array antenna design and the goal of all-in-one arrays for base-stations. Proposed solutions include multiband~\cite{so2013,moradi2012} as well as broadband solutions~\cite{Elsallal2011,Chen2012,Kolitsidas+Jonsson2013}. Comparisons between these classes of arrays are difficult and complex; the evaluation measure may include bandwidth, scan coverage, size and return-loss that have possibly different weight in the evaluation. Similarly, scan-range and bandwidth are critical parameters in design of single-band radar arrays, whereas thickness is critical for e.g. airborne arrays. Optimization that trade e.g. thickness against bandwidth are challenging.

In this letter we propose an {\it array figure of merit}, for the unit-cell of an array antenna over a planar ground-plane that connects bandwidth, return-loss, array thickness and scan-range. The performance of a diverse class of arrays can hence be compared with each other through the array figure of merit. The figure of merit is a number in the interval $[0,1]$ and we propose it as a new goal parameter in the array design. It is straight forward to evaluate the array figure of merit for measured and simulated arrays as well as for a-priori goal-specified antennas. It also provide a first-principle trade-off relation for bandwidth, return-loss, thickness and scan-range. 

The array figure of merit is based on a Rozanov type sum-rule~\cite{Rozanov2000} for waves with linear polarization and its modification to periodic structures~\cite{Sjoberg2009c,Gustafsson+Sjoberg2011}. The key property to obtain the sum-rule here is that the lowest order Floquet-mode reflection coefficient has certain analytic properties in a (generalized) frequency domain due to that the system is causal, linear, passivity and reciprocal~\cite{Bernland+etal2011a}. A bound on array-antenna bandwidth was recently published in~\cite{Doane+etal2013}. The present study extends and widens their result to include multi-band, scan-range, and the properties of the unit-cell structure, thus providing a tighter bound on the array performance. 

Energy and sum-rule based bounds has recently improved the Chu-bound for small antennas see e.g.~\cite{Gustafsson+etal2009a,Yaghjian+Best2005,Vandenbosch2011,Gustafsson+etal2012a,Gustafsson+Jonsson2013}. These small-antenna bounds give essentially tight limits on the possible bandwidth of small antennas given their physical size. For absorbers over a ground plane and for periodic high-impedance surfaces, Rozanov~\cite{Rozanov2000} and~\cite{Gustafsson+Sjoberg2011,Samani+Safian2010} derived bounds on bandwidth. The absorber bound has been throughly investigated see e.g.~\cite{Kazemzadeh+Karlsson2010b}, for a fixed angle of incidence. 

In addition to the above outlined definition of the figure of merit, this letter ends with an investigation of twelve published wideband and/or wide-scan array antennas with respect to our measure. This investigation also empirically gives a first indication on values of the figure of merit that defines an excellent array behavior in the given parameters. Let us clearly note that all the selected arrays are well-designed antennas, meeting high standards, of a wider scope than evaluated here with the array figure of merit. 

\section{Theory}
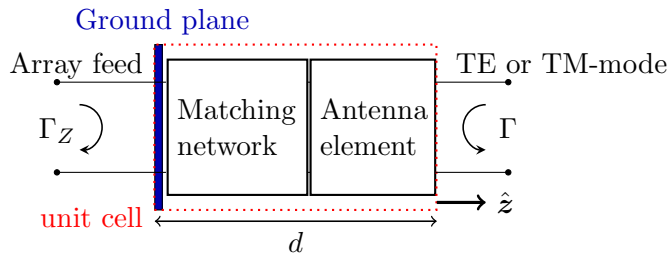
\begin{figure}
\begin{center}
\begin{tikzpicture}
\draw[fill=black!30!blue] (-1.8,-1.1) rectangle (-1.7,1.1) node[above,color=black!30!blue] {Ground plane};
\draw (-.7,0) node [thick, rectangle,draw,text width=1.6cm, minimum height=1.8cm] (bbox) {Matching network};
\draw (1.1,0)node [thick, rectangle,draw,text width=1.4cm,minimum height=1.8cm,label=35:{\ TE or TM-mode}] (abox) {Antenna element};
\draw (.23,.6) rectangle (.27,-.6);
\draw (+.07,0) node [thick, dotted, red, rectangle, draw, minimum height=2.2cm, minimum width=3.75cm,label=206:\color{red}{unit cell},label=165:{Array feed}] {};
\draw [very thick,->] (1.95,-1) -- (2.6,-1) node[right] {$\hz$};
\draw []
(-1.65,.6) -- (-3.1,0.6) 
node[minimum size=2pt,inner sep=0mm,circle,draw,fill=black] {}  node[below=.35cm] {$\Gamma_Z$}
(-1.65,-.6) -- (-3.1,-0.6)
node[minimum size=2pt,inner sep=0mm,circle,draw,fill=black] {}  
(1.93,.6) -- (2.9,0.6)
node[minimum size=2pt,inner sep=0mm,circle,draw,fill=black] {}  node[below=.35cm] {$\Gamma$}
(1.93,-.6) -- (2.9,-0.6)
node[minimum size=2pt,inner sep=0mm,circle,draw,fill=black] {}  
;
%\node (-3,0) {$\Gamma_Z$};
\draw [->,>=stealth,semithick] (-2.8,.3) arc(90:-90:.3cm) ;
\draw [->,>=stealth,semithick] (2.6,.3) arc(90:270:.3cm) ;
\draw [<->,semithick] (-1.8,-1.25) -- node[below] {$d$} (1.95,-1.25); 
\end{tikzpicture}
\caption{Two-port network of the array unit-cell, defining $\Gamma_Z$ and $\Gamma$. The port on the right, is either the fundamental TE or TM Floquet mode. See also~\cite{Doane+etal2013}. Matching in the cell is included in the figure of merit.}\label{figA}
\end{center}
\end{figure}

Consider a passive, infinitely periodic structure of thickness $d$, over a planar ground plane. A linearly polarized incident plane wave of angular frequency, $\omega$, impinges on the periodic structure. It arrives at an angle $\theta$ from the normal of the array. Denote the plane wave fundamental Floquet TE- or TM-mode co-polarized reflection coefficient with $\Gamma$. 
Here the TE(TM)-mode corresponds to that the E(H)-field is orthogonal to the surface normal, $\hz$, see fig.~\ref{figA}. 
Since $\Gamma$ is holomorphic and bounded in magnitude by one in a complex half-plane its logarithm satisfy the sum-rule~\cite{Guillemin1949,Rozanov2000,Bernland+etal2011a,Doane+etal2013}
\begin{equation}\label{1}
I(\theta):=\int_0^\infty \omega^{-2}| \ln\big|\Gamma(\omega,\theta)|\big|\diff \omega \leq q(\theta).
\end{equation}
The term $q$ is $\pi/2$ times the upper bound of the low-frequency linear term of the reflection coefficient, which is a consequence of the sum-rule. 
It is given by~\cite{Gustafsson+Sjoberg2011}:
\begin{equation}\label{q}
q(\theta)=\frac{\pi d}{c}(1 + \frac{\tilde{\gamma}}{2dA})\cos\theta.
\end{equation}
Here $c$ is the speed of light and $A$ is the area of the unit-cell, $d$ is the thickness, and $\tilde{\gamma}$ is a generalized polarizability:
\begin{equation}\label{yy}
\tilde{\gamma}:=\left\{\begin{array}{ll} \gamma_{mt}, & \text{TE} \\
(\gamma_{mt} + \gamma_{ezz}\sin^2\theta)\frac{1}{\cos^2\theta}, & \text{TM}. \end{array}\right.
\end{equation} 
To define $\gamma_{mt},\gamma_{ezz}$, let $\he$ be the unit-vector of the electric field, and $\hht:=\he\times\hz/|\he\times\hz|$. Now, $\gamma_{mt}$ is the projection of the magnetic polarizability tensor in the $\hht$-direction: $\gamma_{mt}=\hht\cdot\boldsymbol{\gamma}_m\cdot \hht$ and the projection of the diagonal electric polarizability is $\gamma_{ezz}=\hz\cdot\boldsymbol{\gamma}_e\cdot\hz$. Note that $\gamma_{mt}$ depends on $\theta$ and polarization through $\hht$. The unit-cell shape and its and materials, i.e. its structure information, enters explicitly in the above inequality through $q$.

To transform the above absorption bound to a bound for the voltage reflection coefficient, $\Gamma_Z$, at the array feed port, a relation between $\Gamma_Z$ and $\Gamma$ is required. Doane {\it et al.}~\cite{Doane+etal2013} showed that $|\Gamma_Z|=|\Gamma|$, when they considered the array unit-cell as a two-port, where the first port is the array feeding port and second is one of the fundamental Floquet modes, e.g. either the TE-mode or the TM-mode, see Fig.~\ref{figA}. This result holds under the assumption that the structure is lossless, reciprocal and without cross-polarization, seen as a two port network, for frequencies below the grating-lobe onset limit. 

Note that the both $I(\theta)$ and $q(\theta)$ depend on the entire unit-cell structure, above the ground plane. A consequence of this is that a matching-network in the unit-cell is automatically included in the array figure of merit. Additional matching below the ground plane is not included here.

The integrand of $I(\theta)$ is positive, hence we can estimate it from below by limit the integration range to $[0,\omega_G(\theta)]$, where $\omega_G(\theta)$ is the onset of the grating lobes. We find that $I_G(\theta)\leq I(\theta)$ now using the relation $|\Gamma|=|\Gamma_Z|$ we find 
\begin{equation}\label{1a}
I_{G}(\theta)=\int_0^{\omega_G(\theta)} \omega^{-2}| \ln\big|\Gamma_Z(\omega,\theta)|\big|\diff \omega 
 \leq q(\theta).
\end{equation}
This is the starting point for our derivation of the array-figure of merit. 

\section{Array figure of merit}
To arrive to the desired array figure of merit, we need to include scan-range and multiple bands. Recall that $q(\theta)$ for the TE-case (or the H-plane) and the TM-case are different. However given $q$ we can obtain the derivation for both cases directly.  Define the scan-range from the array surface normal as $R:=[\theta_0,\theta_{1}]$. We reformulate~\eqref{1a} and note that: 
\begin{equation}\label{2}
\eta_0: = \max_{\theta\in [\theta_0,\theta_1]} \frac{I_G(\theta)}{q(\theta)} \leq 1
\end{equation}

To express $I_G(\theta)$, in terms of bandwidth we need an estimate of the reflection coefficient. There are at least two ways to do this estimate, including e.g. a resonance model. Here we investigate wide-band array antennas which often have a flat response in their operating band(s). Consequently we define the maximal input voltage reflection coefficient over the disjoint angular frequency bands $\{B_m\}_{m=1}^M$: $|\Gamma_{Z,\max_m}|=\max_{\theta\in R;\omega\in B_m}|\Gamma_{Z}|$ where  $B_m:=[\omega_{-,m},\omega_{+,m}]$. We require here that $\max_m \omega_{m,+}\leq \min_{\theta\in R} \omega_G$, which amount to that the operation bands are below the onset of the grating lobe. This requirement is used to preserve the relation $|\Gamma|=|\Gamma_Z|$.

Since $|\Gamma_Z|\leq 1$ we find that $\big|\ln |\Gamma_Z(\omega,\theta)|\big|\geq \big|\ln |\Gamma_{Z,\max_m}|\big|$ in each frequency interval for the given scan range. 
Thus 
\begin{equation}
I_G(\theta)\geq \sum_{m=1}^M \big|\ln |\Gamma_{Z,\max_m}|\big|(\omega_{-,m}^{-1}-\omega_{+,m}^{-1}).
\end{equation} 
This allows us to bound~\eqref{2} as
\begin{equation}
\eta_M:=\frac{c\sum_{m=1}^M \big|\ln |\Gamma_{Z,\max_m}|\big|(\omega_{-,m}^{-1}-\omega_{+,m}^{-1})}{\min_{\theta\in R} q(\theta)} \leq \eta_0 \leq 1.
\end{equation}
Here $q(\theta)$ is known for a given unit-cell. We let the above quantity, $\eta_M$ denote the {\it array figure of merit}. Each unit-cell structural shape, and the choice of polarization influence the value of $\Gamma_Z$ and $q$. Methods to calculate and bound $q$ are given in e.g.~\cite{Sjoberg2009v}. Recall, also that return loss $RL:=-20\log|\Gamma_Z|$.

A less tight bound is obtained if $q$ is replaced with $q_0$, where $q_0$ is any pointwise upper limit of $q$. For the TE-case such an estimate is~\cite{Gustafsson+Sjoberg2011} $q\leq q_0:=\frac{\pi d \mu_s}{c}\cos \theta$, where $\mu_s$ is the maximal value of the static permeability within the unit-cell, and $c$ is the speed of light in vaccum. This results in:
\begin{equation}\label{xx}
\eta_{M}^{TE}:= \frac{c\sum_{m=1}^M \big|\ln |\Gamma_{Z,\max_m}|\big|(\omega_{-,m}^{-1}-\omega_{+,m}^{-1})}{\pi \mu_s d \cos \theta_1} \leq \eta_M \leq 1.
\end{equation}
The above results provide a trade-off between thickness, scan-range and the possible bandwidth of multiple frequency-bands. 

It is easy to re-formulate it as a bound for one of these quantities given the others, but the trade-off between these quantities are often more interesting. We thus introduce $\eta_M^{TE}$ as the TE {\it array figure of merit}, for the multiband case, over the scan range $R$. Given one frequency band, it is clear that if $1/\omega_+$ is small, then \eqref{xx} provides a lower bound on the lowest angular frequency $\omega_-$. We can also rewrite the one frequency band case into the form: 
\begin{equation}\label{4}
\eta^{TE}: = \frac{\big|\ln |\Gamma_{Z,\max}|\big|(BW-1)}{2\pi^2\mu_s (d/\lambda_{\text{hf}}) \cos\theta_{1}}\leq 1.
\end{equation}
Here $1<BW=\omega_{+}/\omega_{-}$ and $\lambda_{\text{hf}}=2\pi c/\omega_{+}$. 
For the special case of $\theta_0=\theta_1$ we obtain the bound in~\cite{Doane+etal2013} in their eqn. (8), this corresponds to study the array figure of merit for a fixed scan angle. 
Eqn. (9) in~\cite{Doane+etal2013} as well as \eqref{xx} above, allow infinite bandwidth for finite thickness arrays.  However both results are based on~\eqref{yy} which is derived for the low-pass case, and furthermore $\omega_+<\omega_G$. It thus remains an open question if infinite bandwidth is possible for finitely thick arrays over a ground plane~\cite{Doane2013}. 

The TM-case is somewhat more complicated and depend also on the value of $\gamma_{ezz}$. For finite conductivity structures, it is shown in~\cite{Gustafsson+Sjoberg2011} that the isotropic slab with $q_0(\theta)=\frac{\pi d}{c}(\frac{1}{\eps_s}\cos\theta+(\mu_s-\frac{1}{\eps_s})\frac{1}{\cos\theta})$, is an upper bound, i.e. $q\leq q_0$, when $\mu_s$ and $\eps_s$ are chosen to their respective maximal static values in the unit-cell domain. PEC structure behavior can be bounded through $\eps_s\rightarrow \infty$ in $q_0$ yielding $q(\theta)\leq \frac{\pi d\mu_s}{c\cos\theta}$, see
also the appendix in~\cite{Gustafsson+Sjoberg2011}. Let's define an equivalent refraction index $n^2=\eps_s\mu_s$, where $\eps_s$ and $\mu_s$ are given their respective maximal values. For $n=1$ we find that $\eta_M^{TM}=\eta_M^{TE}$ and for any $n>1$ we have that $\cos\theta\leq \frac{c}{\pi\mu_s d}q_0(\theta)< \frac{1}{\cos\theta}$. Let $\theta_*$ be the point that minimize the isotropic slab $q_0(\theta)$ in $R$ for the TM-case, then:
\begin{equation}
\eta_{M}^{TM}:= \frac{c\sum_{m=1}^M \big|\ln |\Gamma_{Z,max_m}|\big|(\omega_{-,m}^{-1}-\omega_{+,m}^{-1})}{\pi \mu_s d \big(\frac{1}{n^2}\cos\theta_*+(1-\frac{1}{n^2})\frac{1}{\cos\theta_*}\big)} \leq \eta_0 \leq 1.
\end{equation}
For $n\in [1,\sqrt{2}]$, define $\theta_n:=\arccos(\sqrt{n^2-1})$. We can then specify $\theta_*$ as 
\begin{equation}
\theta_* = \left\{ \begin{array}{ll} \theta_1, & \text{for}\ \theta_1<\theta_n\ \text{and}\ n\in [1,\sqrt{2}], \\
\theta_n, & \theta_n\in [\theta_0,\theta_1]\ \text{and}\ n\in [1,\sqrt{2}], \\ 
\theta_0, & \theta_0>\theta_n\ \text{or}\ n>\sqrt{2}. \end{array}\right.
\end{equation}
For wideband, single band arrays $\eta^{TM}:=\eta^{TM}_1$ reduces to 
\begin{equation}\label{12e}
\eta^{TM} = \frac{\big|\ln |\Gamma_{Z,max}|\big|(BW-1)}{2\pi^2\mu_s (d/\lambda_{\text{hf}}) \big(\frac{1}{n^2}\cos\theta_*+(1-\frac{1}{n^2})\frac{1}{\cos\theta_*}\big)} \leq 1
\end{equation}
For the PEC case with $n\rightarrow \infty$, $q_0(\theta_*)=\frac{\pi\mu_s d}{c \cos\theta_0}$, and if $\theta_0=\theta_1$ we recover the TM-bound in~\cite{Doane+etal2013}. Modification of the figures of merit to other scan-ranges and more complicated combinations of scanning versus frequency band follow directly from the procedure outlined in Eqns.~\eqref{2}-\eqref{12e}.

\section{Investigation of published arrays}

\begin{figure}
\begin{center}
\pgfplotsset{
    standard/.style={
        axis x line=middle,
        axis y line=middle,
        enlarge x limits=0.15,
        enlarge y limits=0.15,
        every axis x label/.style={at={(current axis.right of origin)},anchor=north west},
        every axis y label/.style={at={(current axis.above origin)},anchor=north east}
    }
}
\begin{tikzpicture}
\begin{axis}[standard,xlabel=$d/\lambda_{\text{hf}}$,ylabel=$\eta^{TE}$,
axis x line=bottom,
axis y line=center,
%extra y ticks={0.1,.7},
width=16.0cm,
height=8cm,
grid=major,
ymin=0.1,
ytick={0.1,0.2,...,0.7},
legend pos= south east
]

\addplot+[
%scatter,scatter src=y,
%,mark=*,
%%mark options={blue},
only marks,
point meta=explicit symbolic,
nodes near coords, nodes near coords align={right},
error bars/.cd, 
y dir=both,y explicit
] coordinates {
% H-plane, TE mode
%
(0.228,0.374) [ ] %[\cite{Huss2005}]  %checked
(0.48,0.24) +- (0,0.01) [\cite{Holland2012p}] % checked
%(
(0.408,0.18) +- (0,0.02) [\cite{Infante2010low}]  % checked
(0.5,0.642)[\cite{Elsallal2011}] % bava   
(0.7,0.544)[\cite{DoaneII2013}]
(0.90,0.441426291) +- (0,.04) [\cite{jones2007}]  % checked Munk, CSA center element measured, 45 uncertainty in vswr  LJ ok. 
(0.716,0.401) +- (0,0.02) [ ]% [\cite{Gustafsson2006use}]  % checked self complementary
(2.75,0.6) +- (0,0.15) [\cite{Maloney2011,Friederich2001} ] %fragmented 2
(3.7338,.183) [\cite{Schuneman2001}]
(2.25,.223) [\cite{Schaubert2007}]
(2.85,.2304) [\cite{Stasiowski2008}]
(3.54,.1898) [{ }] %cite{Kindt2010}]
%[\cite{Kindt2010}]
};

%\node[blue] at (axis cs:3.58,.143) {\cite{Schuneman2001}};
%\node[blue,above] at (axis cs:0.48,.44) {\cite{Holland2012p}};
\node[blue,below right] at (axis cs:0.17,.374) {\cite{Huss2005}};
\node[blue,left] at (axis cs:0.72,.4) {\cite{Gustafsson2006use}};
\node[blue] at (axis cs:3.4,.22) {\cite{Kindt2010}};

%\node[blue,left] at (axis cs:0.7,.54) {\cite{Doane2013}};

%\legend{H-plane,E-plane}
\end{axis}
\end{tikzpicture}
\caption{Array figure of merit for a selection of published antennas. All points are for H-plane data. The used published antenn-data are selected to maximize the array figure of merit.}\label{eta}
\end{center}
\end{figure}
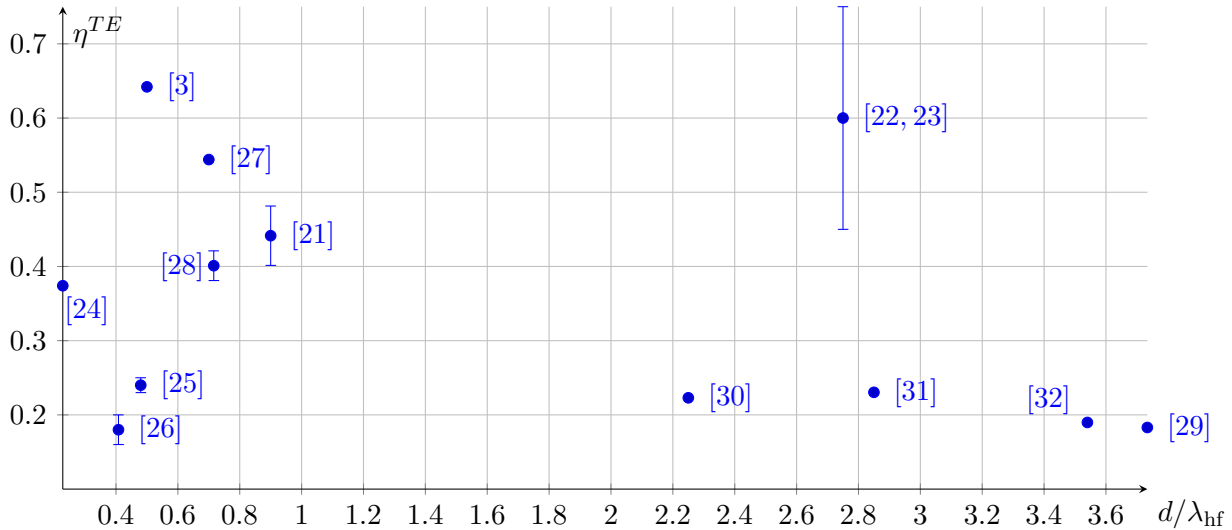

We have selected twelve single wide-band array antennas and/or arrays with large scanning range from broad-side. The selection aims towards finding antennas with a large array figure of merit, but also to investigate different thicknesses and technologies. In representing the array figure of merit we display it as a function of $d/\lambda_{\text{hf}}$. We focused on the H-plane limitation ($\eta^{TE}$), since it is the easier case to calculate. Due to the sparseness of the data in some of the publications, these points are an approximation of their array figure of merit. To illustrate, note that in e.g.~\cite{jones2007} we find that the information on the voltage standing wave ration (VSWR) is only partly given with respect to scanning, which introduces an uncertainty that is indicated with the error-bars in~Fig~\ref{eta}. Similarly in~\cite{Maloney2011,Friederich2001} the value for the return-loss is absent, we have pessimistically used the value of $-5$dB, and the bars indicate $\pm 1$dB. The investigated arrays use non-magnetic materials, and we set the maximal static permeability to $\mu_s=1$. The scan range used for the evaluation are chosen as $[0,\theta_1]$, where $\theta_1$ is determined by the available data and also to maximize the figure of merit. 

To evaluate the array figure of merit, we need to identify two main unknown parameters $BW=\omega_+/\omega_-$ and $|\Gamma_{Z,max}|$. Clearly it is possible to trade $|\Gamma_{Z,max}|$ against bandwidth. The method that we use to identify a high $\eta^{TE}$ is to utilize an ideal rectangular mask between $\omega_-$ and $\omega_+$, with a lower (mask-level) edge on the VSWR($|\Gamma_{Z,max}|$). A well designed antenna may have a fairly flat reference level in the VSWR over the operating bandwidth, above which the rectangular mask (frequency, VSWR($|\Gamma_{Z,max}|$)) can be fitted, such that VSWR($|\Gamma_Z|$) for all $\theta\in R$ and $\omega\in[\omega_-,\omega_+]$ are outside the mask. Note that for antennas with oscillating reflection coefficients/VSWR's in the working band, this approximation method tend to give an underestimate of the figure of merit. 

The investigated arrays shows that the maximal obtained TE-array figure of merit in e.g.~\cite{Holland2012p} occurs for $[0,\theta_1]$, with $\theta_1=\pi/4$, whereas~\cite{Elsallal2011} has essentially the same figure of merit for broadside as for $\theta_1=\pi/4$. The broadside case in~\cite{DoaneII2013} has a slightly higher $\eta^{TE}$ than the case that includes the scan-range. 
Figure of merit considerations shows that  
the tuned resonances separate somewhat and locally push the VSWR level higher faster than $\cos\theta_1$ can compenstate.

The array figure of merit is derived for a lossless unit-cell, but the published array data are, in contrast, often simulated or measured on a finite size array antenna. The effect of the finite size and external (outside the unit-cell) matching network are not included in~Fig.~\ref{eta}. Let us once more repeat that the reviewed arrays are optimized for a given architecture, and a problem dependent specification, and not towards the array figure of merit investigated here. 

Antennas with an oscillating voltage reflection coefficient in its working band may improve their figure of merit with a different type of $|\Gamma_Z|$ estimate. Even for antennas with small oscillations in $|\Gamma_Z|$ for $\omega\in[\omega_-,\omega_+]$ we note that the estimate naturally tend to give an underestimate of $\max_{\theta\in R} I_G(\theta)/q$. 
If we calculate $\max_{\theta\in R} I_G(\theta)/q$ in the provided frequency band in eg.~\cite{Elsallal2011} for the TE-case, we find that $\eta_0 \sim 0.8$, with $q=\frac{\pi\mu_s d}{c}\cos\theta_1$. Hence we conclude that the method of letting  $\max_{\theta\in R,\omega\in B_1}|\Gamma_Z|$  replace $|\Gamma_Z|$ introduces an underestimate of $I_G$ amounting to: $\eta_0-\eta^{TE}$=0.16 for this antenna. The `missing' area of $\max_{\theta\in R} I_G/q$ in the estimate, is largely due to how rapidly $|\Gamma_Z|$ moves from 1 to the working-band level. 

Note that the evaluated array figure of merit also proposes an approximation on the how well the figure of merit predicts performance. The maximum observed value of $\eta^{TE}$ is $0.64$. It remains an open question if the bound can be made tighter while still providing a bandwidth measure.

\section{Conclusion}
A novel {\it array figure of merit} is derived for arrays over a ground-plane. It provides a tool to compare both wide single- and wide multi-band arrays across antenna array technologies. The figure of merit includes bandwidth(s), scan-range, voltage-reflection coefficient and the structural information of the unit cell of the array, which reduce to thickness and material coefficients, in the more generous bound. The array figure of merit also provides a tighter bound than~\cite{Doane+etal2013}, since it allows us to include the unit-cell structure, scan-range and multi-bands. For PEC arrays with a single-band, and a single scan-angle we note that the array figure of merit reduce to the bound in~\cite{Doane+etal2013}. 

We investigated twelve wide-band and/or wide-scan antennas over a large range of thicknesses for the H-plane behavior, all satisfying a figure of merit $\leq 1$, agreeing with the theory. Furthermore, two investigated arrays have a figure of merit $\geq 0.6$, empirically defining them as a high performance arrays in terms of the array figure of merit. 

For most considered antennas, however, the figure of merit is $<0.5$ that indicate one of several possibilities: 1) the reflection coefficient oscillates in the working band and a resonance model could provide a more accurate approximation of their performance, 2) the used technology might not efficiently utilize the thickness, or 3) that the structural unit-cell influence on the array figure of merit may be large. 

The array figure of merit is currently limited to linear polarization, an extension to circular polarization requires another $q$ in \eqref{q}, see~\cite{Gustafsson+Sjoberg2011}. The method is currently limited to infinite arrays over a planar ground-plane, since it is based on a Rozanov-type bound on unit-cells. To extend it to arrays of differently shaped ground-planes remains a research challenge. 

Our array figure of merit provides information about how e.g. bandwidth can be traded against reflection coefficient or antenna thickness. This a-priori information is important in evaluating and designing arrays for wide-band applications. In addition we find that the figure of merit is surprisingly simple to calculate both for measured/simulated antennas as well as for design specifications. Hence, it easily provides an estimate on performance in important antenna parameters. 

\section*{Acknowledgement} 
This work has been supported by Swedish Governmental Agency for Innovation Systems (VINNOVA) within the VINN Excellence Center Chase and NFFP5.


\end{document}